\documentclass[11pt]{article}

\usepackage[english]{babel}
\usepackage[dvips]{epsfig}
\usepackage{epsf}
\usepackage{fullpage}

\newcommand{\mod}{\; {\rm mod} \;}
\newtheorem{theorem}{Theorem}[section]
\newtheorem{lemma}[theorem]{Lemma}
\newtheorem{definition}[theorem]{Definition}
\newtheorem{cor}[theorem]{Corollary}
\newtheorem{rqe}[theorem]{Remark}
\newtheorem{prop}[theorem]{Proposition}
\newtheorem{example}[theorem]{Example}
\newtheorem{fact}{Fact}

\newenvironment{proof}{\par\noindent {\bf Proof.} \rm}{\ ~~~$\fbox{}$}
\newenvironment{proof2}[1]{\par\noindent {\bf Proof of #1.} \rm}{\ ~~~$\fbox{}$}

\newcommand{\sub}[2]{#1{[#2]}}
\newcommand{\card}[1]{{{\rm Card(}#1{\rm)}}}
\newcommand{\TSU}{{T_{A,k}}}
\newcommand{\SPK}{k{\rm PF}(A)}
\newcommand{\TSUA}{{U_{k,A}}}
\newcommand{\TSUC}{{V_{k,A}}}
\newcommand{\TSUB}{\SPK \cap \TSUC}

\newcommand{\fromAtoB}{from $A^*$ to $B^*$}

\begin{document}
\thispagestyle{empty}
\begin{center}
LaRIA~: Laboratoire de Recherche en Informatique d'Amiens\\
Universit\'e de Picardie Jules Verne -- CNRS FRE 2733\\
33, rue Saint Leu, 80039 Amiens cedex 01, France\\
Tel : (+33)[0]3 22 82 88 77\\
Fax : (+33)[0]03 22 82 54 12\\
\underline{http://www.laria.u-picardie.fr}
\end{center}

\vspace{7cm}

\begin{center}
\parbox[t][5.9cm][t]{10cm}
{\center

{\bf Existence of finite test-sets\\ for $k$-power-freeness of uniform morphisms

\medskip

G. Richomme$^{\rm a}$, F. Wlazinski$^{\rm b}$
}
\bigskip

\textbf{L}aRIA \textbf{R}ESEARCH \textbf{R}EPORT~: LRR 2005-10\\
(December 2005)
}
\end{center}

\vfill

\hrule depth 1pt \relax

\medskip

\noindent
$^a$, $^b$ LaRIA, Universit\'e de Picardie Jules Verne, \{gwenael.richomme,francis.wlazinski\}@u-picardie.fr

\vspace{-2cm}
\pagebreak

\title{Existence of finite test-sets\\ for $k$-power-freeness of uniform morphisms\\}
\author{G. Richomme, F. Wlazinski\\
 e-mail: \{gwenael.richomme,francis.wlazinski\}@u-picardie.fr
\\
LaRIA, Universit\'e de Picardie Jules Verne\\ 33 Rue Saint Leu,
80039 Amiens cedex 01, France\\
}
\date{\today}
\maketitle

\begin{abstract}
A challenging problem is to find an algorithm to decide whe­ther a
morphism is $k$-power-free. We provide such an algorithm when $k \geq
3$ for uniform morphisms showing that in such a case, contrarily to
the general case, there exist finite test-sets for $k$-power-freeness.
\end{abstract}


\section{Introduction}

Repetitions in words is a recurrent subject of study in Combinatorics
on Words.  The reader can consult for instance
\cite{CK1997,Lot1983,Lot2002,Lot2005} for surveys of results and
applications.  The interest for such regularities dates back to the
works of A.~Thue \cite{Thu1906,Thu1912} (see also \cite{Ber1992,Ber1995})
who, one century ago, provided examples of some repetition-free words,
more precisely some square-free and overlap-free words.
The construction of some of these words is simple: they are generated as
fixed points of free monoid morphisms.  An example is the fixed point
(denoted $\Theta^\omega(a)$) of the morphism $\Theta$ defined by
$\Theta(a) = abc$, $\Theta(b) = ac$ and $\Theta(c) = b$:
$$\Theta^\omega(a) = abcacbabcbacabcacbacabcb\ldots$$ This word is
$k$-power-free \cite{ISt1977,Thu1912} for any integer $k \geq 2$, that
is, it does not contain any word on the form $u^k$ with $u$ non-empty. May be
strangely, for any $k \geq 2$, the morphism $\Theta$ is not itself
$k$-power-free: it does not map all $k$-power-free words on
$k$-power-free words ($\Theta(ab^{k-1}a) = ab(ca)^kbc$). So where'as
any $k$-power-free morphisms generates a $k$-power-free word, the
converse does not hold.

F.~Mignosi and P.~S\'e\'ebold \cite{MS1993} have proved that it is
decidable whether a morphism generates a $k$-power-free word: more
precisely they proved that, given a word $w$ and a morphism $f$, it is
decidable whether the language $\{f^n(w) \mid n \geq 0\}$ is
$k$-power-free.  However, given an integer $k \geq 3$, to decide if a
morphism is $k$-power-free is still an open problem even if some
partial results have been achieved especially for morphisms acting on
binary alphabets and for 3-power-free morphisms on ternary alphabets
\cite{BEM1979,Ker1986,Lec1985,RW2002,Wla2001}. We note that the case
$k = 2$ was solved by M.~Crochemore \cite{Cro1982}. We also observe
that properties of $k$-power-free morphisms are badly known (see for
instance \cite{RS2004}) despite of some efforts in the eighties
\cite{Ker1986,Lec1985} when relations between morphisms and
variable-length codes (in the sense of \cite{BP1985}) were studied.

A related problem is the study of overlap-free morphisms: an
overlap-free word is a word that does not have any factor of the form
$auaua$ with $a$ a letter and $u$ a word; an overlap-free morphism is
a morphism preserving overlap-freeness. The study of overlap-free
binary morphisms provides ideas of simple tests that can be extended
to other classes of morphisms like $k$-power-free morphisms. For
instance, the monoid of overlap-free binary endomorphisms is finitely
generated. Unfortunately this is no longer true for both larger
alphabets and $k$-power-free morphisms \cite{Cro1982,Ric2003,RW2002}.
Another simple idea is to test overlap-freeness using a finite set of
overlap-free words, called test-set for overlap-freeness
\cite{BS1993,RS1999}. Recently \cite{RW2004} we have shown that, in
the general case, a finite test-set exists for overlap-freeness of
morphisms defined on an alphabet $A$ if and only if $A$ is a binary
alphabet. But if we consider only uniform morphisms (the images of the
letters have all the same length), such test-sets always exist. Note
that the study of uniform overlap-free morphisms is natural since all
overlap-free binary endomorphisms are uniform. Another reason to study
uniform morphisms is provided by Cobham's theorem stating that a word
is automatic if and only if it is the image under a 1-uniform morphism
of a fixed point of a uniform morphism (see for instance
\cite{AS2003}). Finally let us mention that uniform morphisms are
sometimes easier to use to give examples of infinite words with
particular properties, as done for instance in \cite{Och2005} where a finite
test-set is provided for morphisms mapping $\alpha^+$-power-free words
onto $\beta^+$-power-free words when $\alpha$ and $\beta$ are two
rational numbers with $1 \leq \alpha \leq \beta \leq 2$.

We started the study of test-sets for $k$-power-freeness of morphisms
in \cite{RW2002} where we obtained a result similar to the case of
overlap-freeness: for $k \geq 3$, a finite test-set exists for
$k$-power-freeness of morphisms defined on an alphabet $A$ if and only
if $A$ is a binary alphabet.  The purpose of this paper is to complete
this work showing that, as for overlap-freeness, \textit{there always
exist test-sets for $k$-power-freeness of uniform morphisms} (see
Theorem~\ref{TestSetUni}). Up we know, the existence of such test-sets
for uniform morphisms was previously stated only for morphisms
defined on two-letter \cite{Ker1984,Ker1986,Wla2001} or three-letter
alphabets \cite{Lec1985}.

Despite of the similarities
between overlap-freeness and $k$-power-freeness, we would like to
stress many differences between the two studies.
Firstly, we mention that the maximal lengthes of words involved in the
test-sets are different since of course in one case they depend on
the parameter $k$ and not just on the size of the alphabet. More
important is the fact that we introduce a new way to tackle the
decidability of repetition-freeness.

We will only consider test-sets for $k$-power-freeness when $k \geq
3$.  Indeed it is well-known that a uniform morphism is 2-power-free
(that is square-free) if and only if the images of 2-power-free words
of length 3 are 2-power-free: in our terminology this means that the
set of 2-power-free words of length 3 is a test-set for
2-power-freeness of uniform morphisms.  The test-sets we obtain are
not so simple and depend on both the value of $k$ and the cardinality
of $A$.

We present our test-sets, main tools for the proof and the proof
itself in Section~\ref{sectionTestSet}, Section~\ref{detailsProof} and
Section~\ref{section6} respectively.

\section{Notations and main definitions}

We assume the reader is familiar
(if not, see for instance \cite{Lot1983,Lot2002})
with basic notions on
words and morphisms. Let us precise our notations and the main definitions.

 Given a finite set $X$, we denote by $\card{X}$ its cardinality, that
 is, the number of its elements.
 An \textit{alphabet} $A$ is a finite set of symbols called \textit{letters}.
 A \textit{word} over $A$ is a finite sequence of letters
 from $A$.
  Equipped with the concatenation operation, the set $A^*$ of words over $A$
 is a free monoid with the \textit{empty word} $\varepsilon$ as neutral element and $A$ as
 set of generators.
 Given a non-empty
 word $u = a_1\ldots a_n$ with $a_i \in A$, the \textit{length}
 of $u$ denoted by $|u|$ is the integer $n$
 that is the number of letters of $u$.
 By convention, we have $|\varepsilon| = 0$.

 A word $u$ is a \textit{factor} of a word $v$ if there exist
 two (possibly empty) words $p$ and $s$ such that $v = p u s$.
 We also say that $v$ \textit{contains} the word $u$ (as a factor).
 If $p = \varepsilon$, $u$ is a \textit{prefix} of $v$.
 If $s = \varepsilon$, $u$ is a \textit{suffix} of $v$.
 A word $u$ is a \textit{factor} (resp. a
 \textit{prefix}, a \textit{suffix}) of a set of words $X$,
 if $u$ is a factor (resp. a prefix, a suffix) of a word in $X$.

 Let $w$ be a word and let $i, j$ be two integers such that $0 \leq i-1
\leq j \leq |w|$.  We denote by
$\sub{w}{i..j}$ the factor $u$ of $w$ such that there exist two words
$p$ and $s$ with $w = p u s$, $|p| = i-1$, $|p u| = j$.  Note that,
when $j = i - 1$, we have $\sub{w}{i..j} = \varepsilon$.  When $i=j$,
we also denote by $\sub{w}{i}$ the factor $\sub{w}{i..i}$ which is the
$i^{\rm th}$ letter of $w$.

 Given two words $w$ and $u$, we denote by $|w|_u$ the number of different
 words $p$ such that $pu$ is a prefix of $w$.
 For instance, if $w = abaababa$, we have $|w|_a = 5$, $|w|_{aba} = 3$.

Powers of a word are defined inductively by $u^0 = \varepsilon$, and
 for any integer $n \geq 1$, $u^n = u u^{n-1}$: such a word is called
 a \textit{$n$-power} when $n \geq 2$ and $u \neq \varepsilon$.  A word is
 \textit{$k$-power-free} ($k \geq 2$) if it does not contain any
 $k$-power as factor.  A set of $k$-power-free words is said
 \textit{$k$-power-free}.

Let us recall two well-known results of combinatorics on words:

\begin{prop}{\rm \cite{Lot1983}}
\label{Lothaire} Let $A$ be an alphabet and $u,v,w$ three words
over $A$.
If $vu=uw$ and $v \neq \varepsilon$ then there exist two words $r$ and
$s$ over $A$ and an integer $n$ such that $u=r(sr)^{n}$, $v=rs$
and $w=sr$.
\end{prop}

\begin{lemma}{\rm \cite{Ker1986,Lec1985}}
\label{intfact} If a non-empty word $v$ is an internal factor of
$vv$ (that is, if there exist two non-empty words $x$ and $y$ such
that $vv=xvy$) then there exist a non-empty word $t$ and two
integers $i,j \geq 1$ such that $x=t^i$, $y=t^j$ and $v=t^{i+j}$.
\end{lemma}

Let $A, B$ be two alphabets.  A \textit{morphism} $f$ \fromAtoB\ is a
mapping \fromAtoB\ such that for all words $u, v$ over $A$, $f(uv) =
f(u)f(v)$.  When $B$ does not have any importance, we will say that $f$ is a
\textit{morphism on} $A$ or that $f$ is defined on $A$. A morphism on $A$ is
entirely known by the images of the letters of $A$.  When $B=A$, $f$
is called an \textit{endomorphism} (on $A$).
Given an integer $L$, $f$ is \textit{$L$-uniform} if for each letter
$a$ in $A$ we have $|f(a)| = L$.  A morphism $f$ is \textit{uniform}
if it is $L$-uniform for some integer $L \geq 0$.  Given a set $X$ of
words over $A$, and given a morphism $f$ on $A$, we denote by $f(X)$
the set $\{f(w) \mid w \in X\}$.

A morphism $f$ on $A$ is \textit{$k$-power-free} if and only if $f(w)$
is $k$-power-free for all $k$-power-free words $w$ over $A$.  For
instance, the \textit{empty morphism} $\epsilon$ ($\forall a \in A$,
$\epsilon(a) = \varepsilon$) is $k$-power-free.


\section{\label{sectionTestSet}Main result}

Let us recall that in all the rest of this paper $A$ is an alphabet
containing at least two letters and $k \geq 3$ is an integer.

Our main result (Theorem~\ref{TestSetUni}) is the existence of
test-sets for $k$-power-freeness of uniform morphisms whatever is $A$ and $k$:
A \textit{test-set for $k$-power-freeness of uniform morphisms} on $A$ is
a set $T \subseteq A^*$ such that, for any uniform morphism $f$ on $A$,
$f$ is $k$-power-free if and only if $f(T)$ is $k$-power-free.

This existence is provided by the set
$$\TSU = \TSUA \cup (\TSUB)$$
where $\TSUA$, $\SPK$ and $\TSUC$ are
defined as follows:
\begin{itemize}
\item $\TSUA$ is the set of $k$-power-free words over
$A$ of length at most $k+1$,
\item $\SPK$ is the set of all $k$-power-free words over
$A$, and
\item $\TSUC$ is the set of
words over $A$ that can be written $a_0w_1a_1w_2 \ldots
a_{k-1}w_ka_{k}$ where $a_0, a_1, \ldots,a_{k}$ are letters of $A$ and
$w_1, w_2, \ldots, w_{k}$ are words over $A$ verifying $||w_i|-|w_j||
\leq 1$ and $|w_i|_a \leq 1$; $\forall 1 \leq i,j \leq k$ and
$\forall a \in A$.
\end{itemize}
In the previous definition, the inequality $|w_i|_a \leq 1$ means that
any letter of $A$ appears at most once in $w_i$.  In particular, it
follows that $\max\{|w| \mid w \in \TSU \} \leq b_{k,A}$ where
$b_{k,A} = k \times \card{A} +k +1$.

\begin{theorem}
\label{TestSetUni}
$\TSU$ is a test-set for $k$-power-freeness of uniform morphisms
on $A$.
\end{theorem}

An immediate consequence is the following corollary that gives a
simple bound for the length of the words whose images we have to check
to verify the $k$-power-freeness of a morphism:

\begin{cor}
\label{Bound}
A uniform morphism on $A$ is $k$-power-free for an integer $k \geq 3$
if and only if the images by $f$ of all $k$-power-free
words of length at most $k \times \card{A} +k +1$
are $k$-power-free.
\end{cor}

\section{\label{detailsProof}Tools}

In this section we recall or introduce some useful tools. May be the
 reader will read them when needed in the proof of
 Theorem~\ref{TestSetUni}, but we would like to present the novelties
 of our approach (from Section~\ref{sectionDec}).

\subsection{ps-morphisms}

A morphism $f$ is a \textit{ps-morphism} (Ker\"anen \cite{Ker1986}
called it \textit{ps-code}) if\\ \centerline{$f(a) = ps$, and $f(b) = ps'$,
$f(c) = p's$} with $a,b,c \in A$ (possibly $c = b$), $p$, $s$, $s'$,
$p'$ in $B^*$ then necessarily $b = a$ or $c = a$.
Any any ps-morphism is injective. A basic result about these morphisms is:

\begin{lemma}{\rm \cite{Ker1986,Lec1985}}
\label{lemmeSPKPS}
If all the $k$-power-free words of length at
most $k+1$ have a $k$-power-free image by a morphism $f$, then $f$ is a
ps-morphism.
\end{lemma}

\subsection{\label{sectionDec}Decomposition of $k$-powers}

\begin{figure}[ht]
\begin{center}
\epsf{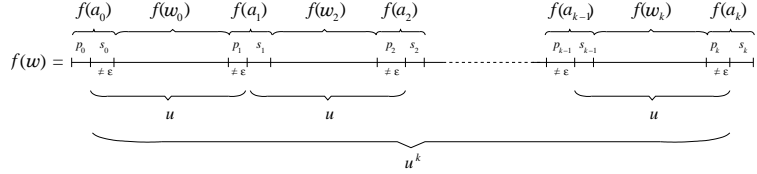}[xscale=5/8,yscale=5/8]
\end{center}
\def\figurename{Figure}
\def\captionseparator{~}
\caption{$(p_i, s_i, x_i, w_i)_{i = 0, \ldots, k}$-decomposition of $u^k$ in $f(w)$\label{decomppk}}
\end{figure}

One situation that we will quickly meet in the proof of
Theorem~\ref{TestSetUni} is: $f$ is a $L$-uniform ps-morphism ($L \geq
0$), $w$ is a $k$-power-free word such that $f(w)$ contains a
$k$-power $u^k$ and $|w| \geq k+1$.  In this case,
Lemma~\ref{lemmeScheme} below will enable us to decompose $u^k$ using factors
of $f(w)$ (see also Figure~\ref{decomppk}).

We observe that (possibly by replacing $w$ by one of its factors)
we can consider that
$u^k$ is {\em directly covered} by $f(w)$. This means that $u^k$ is
not a factor of the image of a proper factor of $w$.
More precisely, if $p_0$ and $s_k$ are the words such that $f(w) =
p_0u^ks_k$ then $|p_0| < L$ and $|s_k| < L$.
The present situation verifies:

\begin{lemma}
\label{lemmeScheme}
Let $f$ be a uniform morphism and let $k \geq 3$ be an integer.  A
$k$-power $u^k$ ($u \neq \varepsilon$) is directly covered by the
image of a word $w$ of length at least $k+1$ if and only if there exist
words $(p_i)_{i = 0, \ldots, k}$, $(s_i)_{i = 0, \ldots, k}$, $(w_i)_{i = 1, \ldots, k}$ and letters $(a_i)_{i = 0, \ldots, k}$ such that:\\
\begin{tabular}{lrcl}
(1) & $w$ & $=$ & $a_0w_1a_1 \ldots a_{k-1}w_ka_k$,\\
(2) & $f(a_i)$ & $=$ & $p_is_i$ ($0 \leq i \leq k$),\\
(3) & $s_0$ & $\neq$ & $\varepsilon$, \\
(4) & $p_i$ & $\neq$ & $\varepsilon$ ($1 \leq i \leq k$),\\
(5) & $u$ & $=$ & $s_{i-1}f(w_i)p_i$ ($1 \leq i \leq k$).\\
\end{tabular}
\end{lemma}

\begin{proof2}{Lemma~\ref{lemmeScheme}}
By definition $f$ is uniform: Let $L$ be the integer such that
$|f(b)| = L$ for each letter $b$.
Clearly if words $(p_i)_{i = 0, \ldots, k}$, $(s_i)_{i = 0, \ldots, k}$,
$(w_i)_{i = 1, \ldots, k}$ and letters $(a_i)_{i = 0, \ldots, k}$ verify
Conditions~(1) to (5), then $|w| \geq k+1$ and $u^k$ is directly covered by
$f(w)$.

Assume now that $u^k$ is covered by $f(w)$ with $|w| \geq k+1$.  Let
$p_0$ and $s_k$ be the words such that $f(w) = p_0u^ks_k$. For each
integer $\ell$ between 0 and $k$, let $i_\ell$ be the least non-zero integer
such that $pu^\ell$ is a prefix of $f(w[1..i_\ell])$.  Since $u^k$ is
covered by $f(w)$, $i_0 = 1$, $i_k = |w|$ and $i_0 \leq i_1 \leq i_2
\leq \ldots \leq i_k$.  If $i_\ell = i_{\ell+1}$ for (at least) one integer
$\ell$ between 0 and $k-1$, then $|u| \leq |f(a_\ell)| = L$.  For any
integer $m$ between 0 and $k-1$, since $f$ is $L$-uniform and $|u|
\leq L = |f(a_m)|$, $i_m +1  \geq i_{m+1}$ ($i_m = i_{m+1}$ or $i_m + 1
= i_{m+1}$).  Hence $|w| = i_k \leq i_{\ell+1} + (k-\ell-1) = i_\ell + (k-\ell-1)
\leq (i_0 + \ell) + (k-\ell-1) = k$: a contradiction.

So $i_0 < i_1 < i_2 < \ldots < i_k$.  We define for each integer
$\ell$ between 1 and $k$ the words $w_\ell = w[i_{\ell-1}+1\ldots
i_\ell-1]$ and $p_\ell$ such that $f(w[1..i_{\ell+1}-1])p_\ell =
p_0u^\ell$.  Moreover let $a_\ell = \sub{w}{i_\ell}$ for $0 \leq \ell
\leq k$.  By construction for $0 \leq \ell \leq k-1$, the word
$p_\ell$ is a non-empty prefix of $f(a_\ell)$ and so we can consider
the word $s_\ell$ such that $f(a_\ell) = p_\ell s_\ell$. Up to now by
construction, we have Conditions~(1), (2), (4) and (5). Since $u^k$ is
covered by $f(w)$, Condition~(3) is also verified.
\end{proof2}

\medskip

\begin{definition}
\label{defDec}\rm
When a $k$-power $u^k$ is directly covered by the image (by a uniform
morphism $f$) of a word of length at least $k+1$, if $(p_i, s_i, a_i,
w_i)_{i = 0, \ldots, k}$ is a $(4k+4)$-uple such that $w_0=
\varepsilon$ and the other $4k+3$ words verify Conditions~(1) to (5)
of Lemma~\ref{lemmeScheme}, we will say that $u^k$ has a $(p_i, s_i,
a_i, w_i)_{i = 0, \ldots, k}$-decomposition in $f(w)$, or that $(p_i,
s_i, a_i, w_i)_{i = 0, \ldots, k}$ is a decomposition of $u^k$ in
$f(w)$.
\end{definition}

\subsection{\label{section4}Non-synchronized decompositions of $k$-powers}

Between all decompositions that a $k$-power can have in the image of a
word by a $L$-uniform morphism $f$, Lemma~\ref{lemmaSynchro} will
allow us to eliminate the following possibility:

\begin{definition}
\label{defSync} Let $(p_i,
s_i, a_i, w_i)_{i = 0, \ldots, k}$ be as in Definition~\ref{defDec}.
When $|s_i| = |s_{i+1}|$ for an integer $i$ between 1 and $k-2$,
the decomposition is said synchronized
(with respect to images of factor of $w$), or shortly that
the $k$-power $u^k$ is \textit{synchronized} in $f(w)$.
\end{definition}

Let us make several remarks about this definition.

First it is immediate that a decomposition $(p_i, s_i, a_i, w_i)_{i =
0, \ldots, k}$ of a $k$-power is synchronized if and only if for all
integers $i, j$ with $1 \leq i < j \leq k-1$, we have $|s_i| = |s_j|$.
Since $f$ is uniform, and since  $f(a_\ell) = p_\ell s_\ell$ (for
all $\ell$, $1 \leq \ell \leq k-1$), it is also equivalent that $|p_i| =
|p_j|$ for all $1 \leq i < j \leq k-1$, or that $|p_i| = |p_{i+1}|$
for all $1 \leq i \leq k-2$.

One aspect may appear strange: why do not we allow $i = 0$ in the
definition of a synchronized decomposition? This is due to the
dissymmetry brought by Conditions~(3) and (4) in the definition of a
decomposition. Assume that $(p_i, s_i, a_i, w_i)_{i = 0, \ldots, k}$
is a synchronized decomposition of a $k$-power $u^k$ in $f(w)$ with
$f$ $L$-uniform. Since $|s_1| = |s_2|$, we have $|p_1| = |p_2|$.
Moreover $u = s_0 f(w_1) p_1 = s_1 f(w_2) p_2$. Thus $p_1 = p_2$ and
$s_0 f(w_1) = s_1 f(w_2)$. When $s_1 \neq \varepsilon$, since also
$p_1 \neq \varepsilon$, we have $0 < |s_1| < L$. In this case $s_0 =
s_1$.  But when $s_1 = \varepsilon$, since $s_0 \neq \varepsilon$, we
have $s_0 \neq s_1$, $p_0 = \varepsilon$ and $s_0 = f(a_0) \neq s_1$.

Of course we do not consider $i = k-1$ in the definition
of a synchronized decomposition simply because $s_k$ is not a factor
of $u^k$.

\begin{lemma}
\label{lemmaSynchro}
Let $f$ be a uniform ps-morphism defined on an alphabet $A$, and let
$k \geq 3$ be an integer.  Any $k$-power directly covered by the image
by $f$ of a $k$-power-free word of length at least $k+1$
is not synchronized.
\end{lemma}

\begin{proof2}{Lemma~\ref{lemmaSynchro}}
By definition $f$ is uniform: Let $L$ be the integer such that
$|f(b)| = L$ for each letter $b$.
Assume there exists a $k$-power $u^k$ that has a
synchronized decomposition $(p_i, s_i, a_i,
w_i)_{i = 0, \ldots, k}$ in $f(w)$, where $w$ is a
$k$-power-free word.  By hypothesis $s_i = s_j$ and $p_i = p_j$ for
all $0 < i < j < k$.  We denote $s = s_1$ and $p = p_1$.  From $u =
sf(w_2)p = sf(w_k)p_k$, we deduce that $|p| = |p_k| \mod L$.  Since
$0 < |p|,|p_k| \leq L$, we get $p_k = p$.  Hence
$u = s_0 f(w_1) p$ and $u =s f(w_i)p$ for all $2 \leq i \leq k$. We
have seen before the lemma's statement that $s_0 = s$ when $s \neq
\varepsilon$ and $s_0 = f(a_0)$ when $s = \varepsilon$.

Assume first $s = \varepsilon$ and $s_0 = f(a_0)$.  Since $f$ is
injective, we get $a_0 w_1a_1 = w_ia_i$ for all
$2 \leq i \leq k$. Thus $w = (a_0 w_1a_1)^k$. This contradicts the
fact that $w $ is $k$-power-free.

So $s \neq \varepsilon$ and $s_0 = s$. Since $f$ is injective, $w_i =
w_1$ for all $1 \leq i \leq k$ and $a_i = a_1$ for all $1 \leq i \leq
k-1$. Hence $w = a_0(w_1a_1)^{k-1}w_1a_k$. Since $w$ is $k$-power-free,
$a_0 \neq a_1$ and $a_k \neq a_1$. Let $a = a_1$, $b = a_k$, $c =
a_0$, $p' = p_0$ and $s' = s_k$: $f(c) = p's$, $f(b)
= ps'$. From $f(a) = ps$, we deduce that $f$ is not a ps-morphism.
\end{proof2}

\medskip

We end this section with some examples of non-synchronized $k$-powers.

\begin{example}
\label{example1}
$f(a) = baaba$, $f(b) = bcdab$, $f(c) =
cdabc$, $f(d) = dbaab$~:
$$f(abcd) = baab (abcd)^3 baab.$$ The decomposition of $(abcd)^3$ in
$f(abcd)$ is given by $a_0 = a$, $a_1 = b$, $a_2 = c$, $a_3 = d$, $w_1
= w_2= w_3 = \varepsilon$, $p_0 = baab = s_3$, $s_0 = a$, $p_1 = bcd$,
$s_1 = ab$, $p_2 = cd$, $s_2 = abc$, $p_3 = d$.
\end{example}

\begin{figure}[ht]
\hspace*{\fill}
\epsf{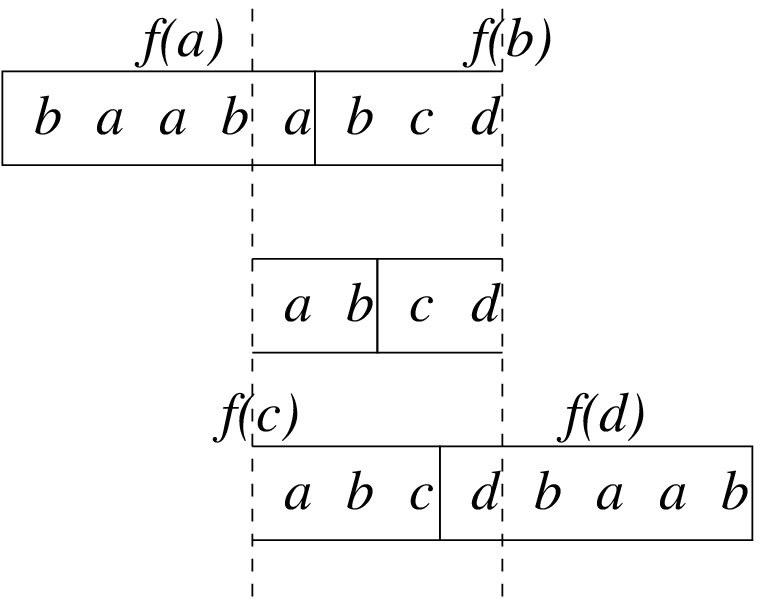}[xscale=1/2,yscale=1/2]
\hfill
\epsf{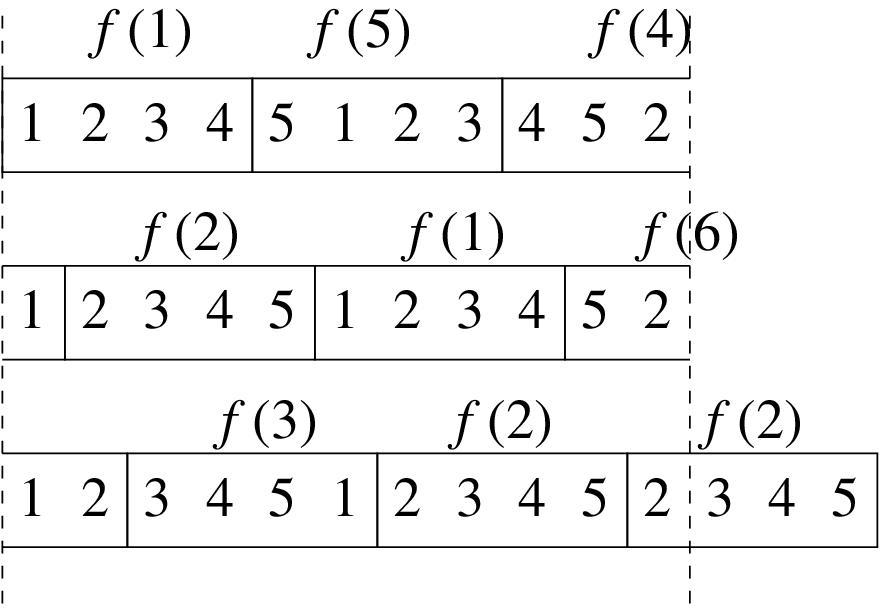}[xscale=1/2,yscale=1/2]
\hspace*{\fill}\\
\hspace*{\fill}
\parbox[t]{0.3\textwidth}{\caption{Example~\ref{example1}}}
\hfill
\parbox[t]{0.3\textwidth}{\caption{Example~\ref{example2}}}
\hspace*{\fill}
\end{figure}

\begin{example}
\label{example2}
$f(1) = 1234$; $f(2) = 2345$, $f(3) = 3451$, $f(4) = 4521$,
$f(5) = 5123$, $f(6) = 5212$~:
$$f(154216322) = (12345123452)^3345$$
The decomposition of $(12345123452)^3$ in $f(154216322)$ is given
by $a_0 = 1$, $a_1 = 4$, $a_2 = 6$, $a_3 = 2$,
$w_1 = 5$, $w_2 = 21$, $w_3 = 32$,
$p_0 = \varepsilon$, $p_1 = 452$, $p_2 = 52$, $p_3 = 2$,
$s_0 = f(a_0)$, $s_1 = 1$, $s_2 = 12$, $s_3 = 345$.
\end{example}


\subsection{\label{section5}Reduction of a $k$-power}

In this section, we introduce the key technic of the proof of
Theorem~\ref{TestSetUni}. It consists in the possibility to reduce the
length of $k$-powers in order to consider only $k$-powers covered by the
image of a word in $\TSUC$.

\begin{prop}
\label{propReduc}
Let $f$ be an injective uniform morphism on $A$.  If there exists a
$k$-power-free word $W$ of length at least $k+1$ such that $U^k$ is
directly covered by $f(W)$ then there exists a word $w$ of length at
least $k+1$ such that $w \in \TSUC$, $|w| \leq |W|$ and $f(w)$ covers
a $k$-power $u^k$. Moreover the $k$-powers $u^k$ and $U^k$ are both
synchronized or both non-synchronized.
\end{prop}

\begin{figure}[ht]
\begin{center}
\ \\[0pt]
\epsfig{file=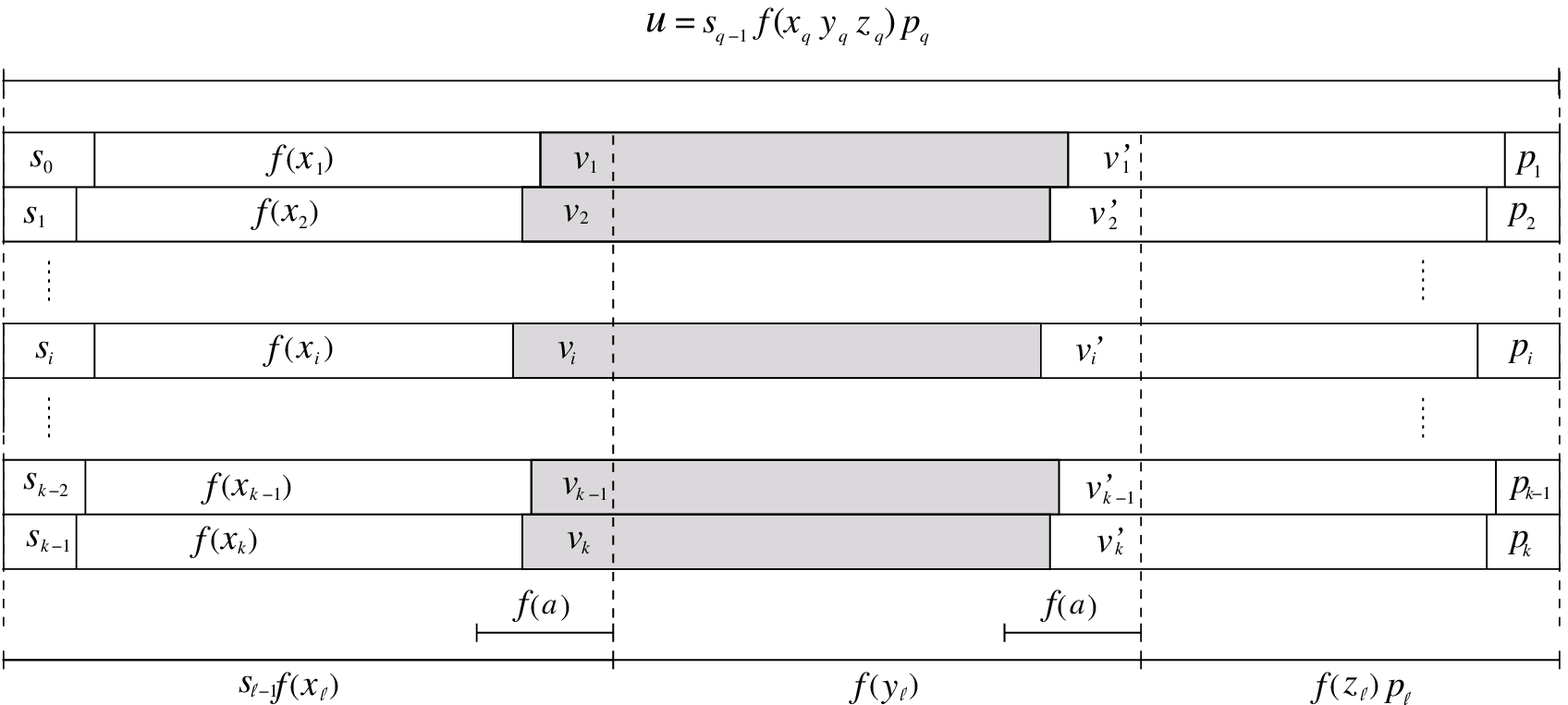,width=\linewidth}
\end{center}
\def\figurename{Figure}
\def\captionseparator{~}
\caption{\label{ukb}}
\end{figure}

This proposition is a direct corollary of Lemma~\ref{reduc}
(to be used inductively) whose idea is illustrated by Figure~\ref{ukb}.

We denote by ${\rm Reduced}(U^k, W)$ the set of pairs $(u^k,w)$ that
can be obtained in conclusion of Proposition~\ref{propReduc}.

\begin{lemma}[Reduction lemma]
\label{reduc}

Let $f$ be an injective uniform morphism on $A$ and let $w$ be a word
over $A$.  We assume that there exists a non-empty word $u$ such that
the $k$-power $u^k$ has a $(p_i,s_i,a_i,w_i)_{i=0,..,k}$-decomposition
in $f(w)$.  We also assume that there exist an integer $1 \leq \ell
\leq k$ and a letter $a$ in $A$ such that
$w_{\ell}=x_{\ell}y_{\ell}z_{\ell}$ and both $x_{\ell}$ and $y_{\ell}$
end with $a$.  Then:
\begin{enumerate}
\item For all integers $i$ such that $1 \leq i \leq k$, there exist three words
$x_i,y_i,z_i$ such that
$w_i=x_i y_i z_i$,
$|s_{\ell -1}f(x_{\ell})|-|f(a)| < |s_{i-1}f(x_i)| \leq |s_{\ell -1}f(x_{\ell})|$ and
$|y_i|=|y_{\ell}|$.
\item Let $u'=s_{\ell -1}f(x_{\ell}z_{\ell})p_{\ell}$ and $w' = a_0
\prod _{i=1}^{k}(x_iz_ia_i)$. The $k$-power $(u')^k$ has a
$(p_i,s_i,a_i,x_iz_i)_{i=0,..,k}$-decomposition in $f(w')$.
\item $|w'| < |w|$.
\end{enumerate}
\end{lemma}

To explain Figure~\ref{ukb}, let us say that the grey parts are
deleted and that the two occurrences of $f(a)$ allow to merge the left
and right non-grey parts in order to have the new $k$-power $(u')^k$
directly covered by the image of the new word $w'$.

\medskip

\begin{proof2}{lemma~\ref{reduc}}

1. By definition $f$ is uniform: Let $L$ be the integer such that
$|f(b)| = L$ for each letter $b$.

Let $i$ be an integer such that $1 \leq i \leq k$.  We have $u =
s_{i-1}f(w_i)p_i = s_{\ell -1}f(x_{\ell}y_{\ell}z_{\ell})p_{\ell}$.
Let us observe that:
$$|s_{i-1}| \leq |s_{\ell-1}f(x_{\ell})| \leq |s_{i-1}f(w_i)|$$
Indeed, since $|f(x_\ell)| \neq 0$ ($x_\ell$ ends with $a$), we have
$|s_{i-1}| \leq |f(a_{i-1})| = L = |f(x_\ell)| \leq
|s_{\ell-1}f(x_\ell)|$.  Moreover 
$|s_{\ell-1}f(x_{\ell})| \leq |s_{\ell-1}f(x_{\ell})|+|f(y_{\ell})|-|f(a)|
\leq |s_{\ell -1}f(w_{\ell})|-|f(a)| \leq |s_{\ell -1}f(w_{\ell})|+|p_{\ell}|-|p_i|
= |s_{i-1}f(w_i)|$.

Thus we can define $x_i$ as the greatest prefix
(maybe empty) of $w_i$ such that $s_{i-1}f(x_i)$ is a prefix of
$s_{\ell -1}f(x_{\ell})$. Since $f$ is uniform, we have:
$$|s_{\ell -1}f(x_{\ell})| - |f(a)| < |s_{i-1}f(x_i)| \leq |s_{\ell -1}f(x_{\ell})|$$

It follows that $|s_{i-1}f(x_i)| \leq |s_{\ell-1}f(x_{\ell})| < |s_{\ell -1}f(x_{\ell}y_{\ell})|$.
Let $y_i$ be
the greatest word such that $x_iy_i$ is a prefix of $w_i$ and
$s_{i-1}f(x_iy_i)$ is a prefix of $s_{\ell-1}f(x_\ell y_\ell)$.
Let $z_i$ be the word such that $w_i=x_iy_iz_i$

Let $v_i'$ be the word such that $s_{i-1}f(x_iy_i)v_i' = s_{\ell
-1}f(x_{\ell}y_{\ell})$.  We have $v_i'f(z_{\ell})p_{\ell}=f(z_i)p_i$.  Assume
$|v_i'| \geq L$. The definition of $y_i$ implies that $z_i=\varepsilon$.  The
equality $|v_i'f(z_{\ell})p_{\ell}|=|p_i|$ with $|p_{\ell}|\neq 0$ is incompatible
with $|p_i| \leq L$. Thus $|v_i'| < L$.  It follows: $$|s_{\ell
-1}f(x_{\ell}y_{\ell})| - |f(a)| < |s_{i-1}f(x_iy_i)| \leq |s_{\ell
-1}f(x_{\ell}y_{\ell})|$$

From this double inequality and the previous one concerning $|s_{i-1}f(x_i)|$,
we deduce that $|f(y_{\ell})| - |f(a)| < |f(y_i)| < |f(y_{\ell})| + |f(a)|$.

Since $f$ is uniform, it follows that $|f(y_{\ell})| = |f(y_i)|$
and $|y_i|=|y_{\ell}|$ (see Figure~\ref{ukb}).


2.  For  all integers $1 \leq i \leq k$,
let $v_i$ be the word such that $s_{i-1}f(x_i)v_i = s_{\ell -1}f(x_{\ell})$.
By definition of $x_i$, we have $0 \leq |v_i|<|f(a)|$.
Moreover $f(y_iz_i)p_i=v_if(y_{\ell}z_{\ell})p_{\ell}$.
Since $|y_i|=|y_{\ell}|$, we get
$|s_{\ell -1}f(x_{\ell})|= |s_{\ell -1}f(x_{\ell}y_{\ell})|-|f(y_{\ell})|$
$=|s_{i-1}f(x_iy_i)v_i'|-|f(y_i)|=|s_{i-1}f(x_i)v_i'|$.
It follows that $|v_i|=|v_i'|$.
Since $x_{\ell}$ and $y_{\ell}$ both end with $a$ and since
$|v_i| = |v_i'| < |f(a)|$, it follows that $v_i$ and $v_i'$ are both
suffixes of $f(a)$ and so $v_i=v_i'$.

Let $w'=a_0 \prod _{i=1}^{k}(x_iz_ia_i)$. For all integers $i$ such that $1 \leq i \leq k$,
we have $u'= s_{\ell -1}f(x_{\ell}z_{\ell})p_{\ell}=s_{i-1}f(x_i)v_if(z_{\ell})p_{\ell} = s_{i-1}f(x_iz_i)p_i$.
Thus, $f(w')=p_0s_0 \prod _{i=1}^{k} f(x_iz_i)p_i s_i$
$= p_0 \left( \prod _{i=1}^{k} s_{i-1}f(x_iz_i)p_i \right) s_k$
$= p_0 u'^ks_k$.

3. Since $y_{\ell} \neq \varepsilon$ ($y_{\ell}$ ends with $a$), we have $|w'|<|w|$.

\end{proof2}

\begin{figure}[ht]
\begin{center}
\epsf{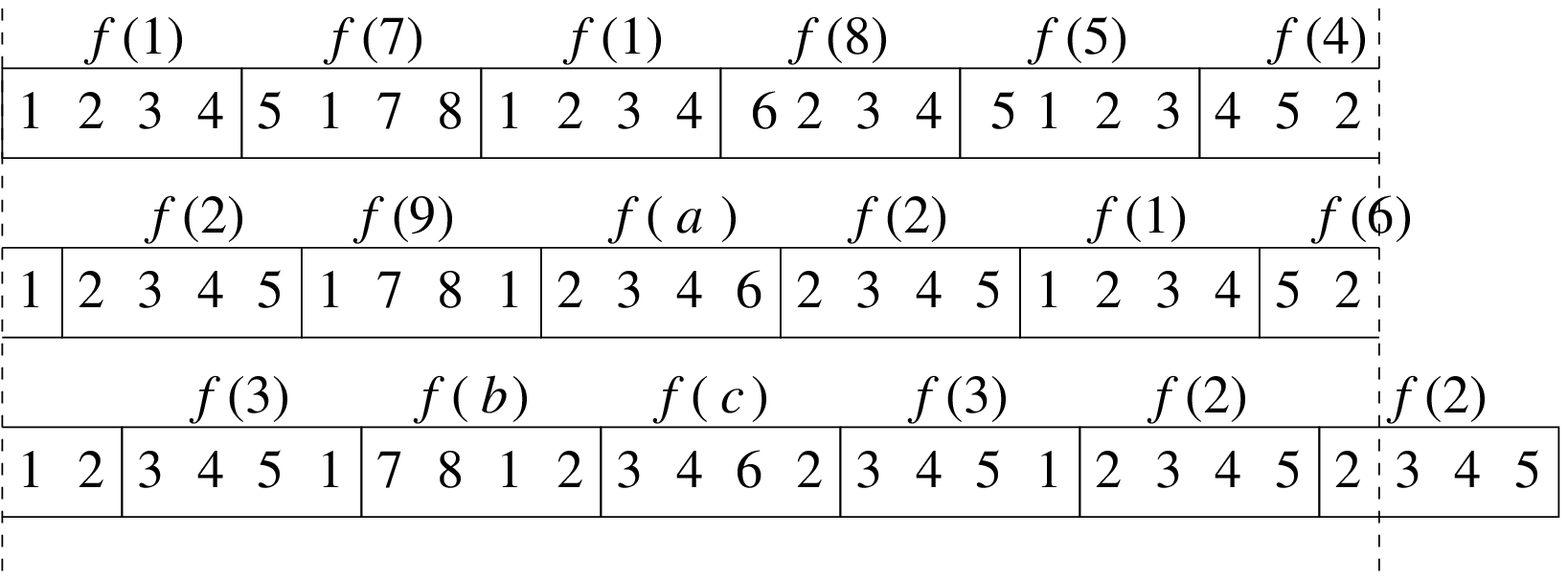}[xscale=1/2,yscale=1/2]
\end{center}
\vspace{-0.5cm}
\def\figurename{Figure}
\def\captionseparator{~}
\caption{\label{fig3.1}Example~\ref{example3}}
\end{figure}

\begin{figure}[ht]
\hspace*{\fill}
\epsf{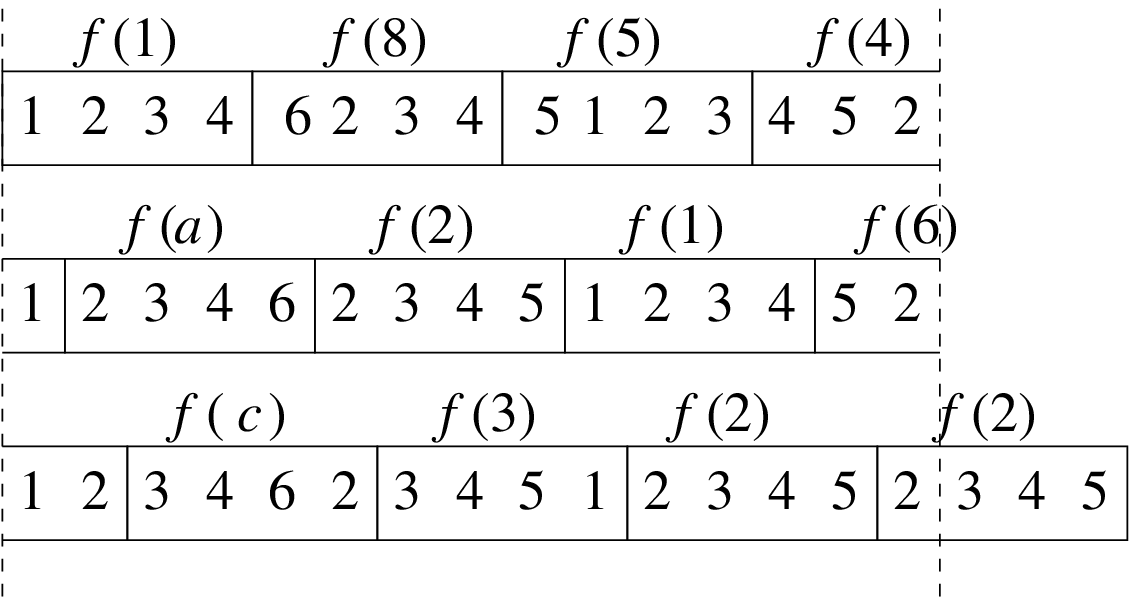}[xscale=1/2,yscale=1/2]
\hfill
\epsf{noSync2.eps}[xscale=1/2,yscale=1/2]
\hspace*{\fill}\\
\hspace*{\fill}
\parbox[t]{0.3\textwidth}{\caption{\label{fig3.2}first possible reduction}}
\hfill
\parbox[t]{0.3\textwidth}{\caption{\label{fig3.3}second possible reduction}}
\hspace*{\fill}
\end{figure}

\medskip
Let us give an example of reduction:

\begin{example}
\label{example3}\rm
Let us consider the morphism defined by $f(1) = 1234$; $f(2) = 2345$,
$f(3) = 3451$, $f(4) = 4521$, $f(5) = 5123$, $f(6) = 5212$, $f(7) =
5178$, $f(8) = 6234$, $f(9) = 1781$, $f(a) = 2346$, $f(b) = 7812$,
$f(c) = 3462$.  This morphism is not $3$-power-free (it is not a
ps-morphism).  We observe (see Figure~\ref{fig3.1}) that
$f(17185429a2163bc322)$ contains the cube
$(12345178123462345123452)^3$. This $3$-power can be reduced on two
ways. First, using the fact that $f(1)$ appears twice in the first
occurrence of $u$, we can obtain the cube $(123462345123452)^3$ in the
image of $f(1854a216c322)$ as shown by Figure~\ref{fig3.2}. Second,
using the fact that $f(3)$ appears twice in the first occurrence of
$u$, we can obtain the cube $(12345123452)^3$ in the image of
$f(154216322)$ as shown by Figure~\ref{fig3.3}.
\end{example}

\bigskip

We observe in Example~\ref{example3} that the two possible reductions
verify the first Reduction Rule, and the different words obtained 
are both in $\TSUC$. The one chosen will be the first
reduction according to the second rule.

\bigskip

We end with two remarks (using notations from Lemma~\ref{reduc}) that
will be useful in the end of Theorem~\ref{TestSetUni}. The first
remark is a direct consequence of the first part of
Lemma~\ref{reduc}. To understand the second remark, we observe that
since $|u|=|s_{\ell -1}f(x_{\ell}y_{\ell}z_{\ell})p_{\ell}|=|s_{j
-1}f(x_jy_jz_j)p_j|$ for all integers $j$ such that $1 \leq j \leq k$,
we also have $|f(z_j)p_j|-|f(a)| < |f(z_{\ell})p_{\ell}| \leq
|f(z_j)p_j|$.

\begin{rqe}\label{spere}~\\
\begin{enumerate}
\item \label{spered3} If there exists an integer $q$ such that $x_q = \varepsilon$
then $x_{\ell}=a$ and $|s_{\ell-1}| < |s_{q-1}|$.
\item \label{spered5}
If there exists an integer $q$ such that $z_q = \varepsilon$ then
$z_{\ell}=\varepsilon$ and $|p_{\ell}| \leq |p_q|$.
\end{enumerate}
\end{rqe}

\subsection{\label{more}More precisions on the reduction}

Proposition~\ref{propReduc} will enable us to prove Theorem~
\ref{TestSetUni} when $k \geq 4$. 

More precisely given two words $W$ and $U$ with $W$ $k$-power-free,
$|W| \geq k+1$ and $U^k$ directly covered by $f(W)$, we will construct
(using this proposition) some words $w$ and $u$ such that $(u^k, w)$
belongs to $ {\rm Reduced}(U^k, W)$, $|w| \leq |W|$ and $w \in \TSUC$.
Moreover the decomposition of $u^k$ in $f(w)$ will be
non-synchronized. Since the word $w$ belongs to $\TSUC$ and since
$f(\TSU)$ is $k$-power-free, we can see that $w$ is not
$k$-power-free, and so there exists a non-empty word $v$ such that
$v^k$ is a factor of $w$. We will be able to prove that this situation
will be possible only if $k = 3$ and $|v| = 1$.

\medskip

But when $k = 3$, the following example shows that there can exists
words $w$ and $u$ such that $u^k$ has a non-synchronized decomposition
in $f(w)$: so we will need to be more precise in our use of the
reductions.

\begin{example}
\label{examplek=3}
\rm
Let $f$ be the morphism from $\{1, 2, 3, 4, 5, 6, 7, 8, 9\}^*$ to
$\{a, 0, 1, 2, 3, 4,$ $5, 6, 7, 8, 9, b\}^*$ defined by
$f(1) = a0123$,
$f(2) = 40125$,
$f(3) = 67892$,
$f(4) = 34012$,
$f(5) = 56789$,
$f(6) = 23401$,
$f(7) = 25678$,
$f(8) = 92340$,
$f(9) = 1234b$. We have (see Figure~\ref{figExK=3}):
$$f(1234445666789) = a(012340125678923401234)^3b$$

Thus this 5-uniform morphism $f$ is a ps-morphism for which there
exists a non-synchronized $k$-power. We let the reader
verify that $f$ is a
3-power-free morphism and so $f(T_{3,\{1,2,3,4,5,6,7,8,9\}})$ is
3-power-free.

\begin{figure}[ht]
\begin{center}
\epsf{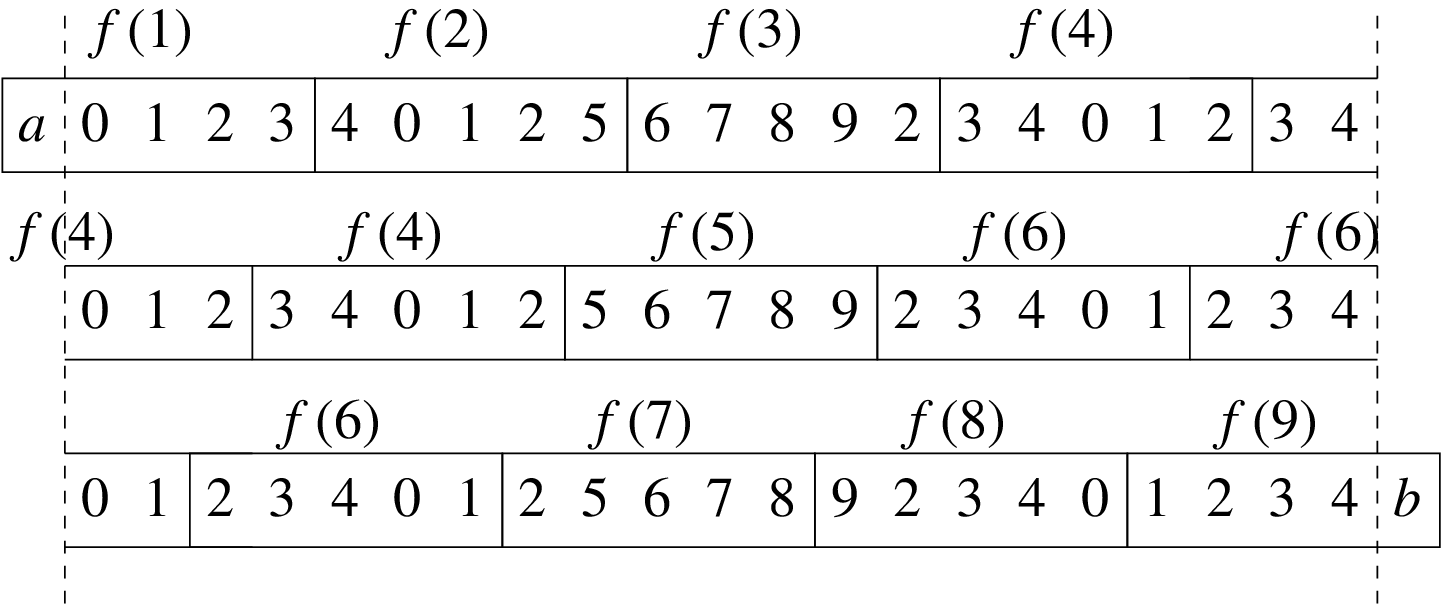}[xscale=1/2,yscale=1/2]
\end{center}
\def\figurename{Figure}
\def\captionseparator{~}
\caption{An example of non-synchronized 3-power\label{figExK=3}}
\end{figure}
\end{example}

We now explain how we tackle the situation $k = 3$ and $|v| = 1$.  As
we have just seen by Example~\ref{figExK=3}, there can exist words $w
\in \TSUC$ and $u$ such that $u^k$ has a non-synchronized
decomposition in $f(w)$.  We will show that, under all current
hypotheses, $w$ and $u$ cannot be obtain by successive reductions from
the words $W$ and $U$ define in the previous section. For this
purpose, we will be more precise on the way the reductions are made to
obtain a couple $(u^k,w)$ in ${\rm Reduced}(U^k, W)$. Actually one can
observe that if a word does not belong to $\TSUC$, there can exist
many different ways to reduce it using Lemma~\ref{reduc}. We will
apply the two following additional rules (with the notations of
Lemma~\ref{reduc}):

\pagebreak

\noindent
\textbf{Reduction rules}:
\begin{enumerate}
\item \label{Newspered2} $|x_{\ell}|_a=1$ and $|y_{\ell}|_a=1$
\item if there exist an integer $1 \leq \ell' \leq k$ and a letter
$a'$ in $A$ such that $w_{\ell'}=x_{\ell'}y_{\ell'}z_{\ell'}$ and both
$x_{\ell'}$ and $y_{\ell'}$ end with $a'$ and such that $(\ell,a) \neq
(\ell',a')$, then $|s_{\ell-1}f(x_\ell)| < |s_{\ell'-1}f(x_{\ell'})|$.
\end{enumerate}

These rules mean that we always made the leftmost reduction
possible. The determinism introduced by these rules will be 
a key element of the proof.

\section{\label{section6}Proof of Theorem~\ref{TestSetUni}}

In this section, we prove Theorem~\ref{TestSetUni} which means:
given any $L$-uniform morphism $f$ on
$A$ (with $L \geq 0$ an integer), $f$ is $k$-power-free if and only
$f(\TSU)$ is $k$-power-free. 

Let $f$ is a uniform morphism from $A^*$ to $B^*$ where $B$ is an
alphabet not necessarily equals to $A$, and let $L$ be the integer
such that $|f(b)| = L$ for each letter $b$.  The ``only if'' part of
the theorem follows immediately from the definition and the
``if'' part is also immediate when $L = 0$. Thus from now on $L \geq
1$. We assume that $f(\TSU)$ is $k$-power-free and we show (by
contradiction) that $f$ is $k$-power-free.

\medskip

Since $\TSUA \subseteq \TSU$, by
Lemma~\ref{lemmeSPKPS}, we have

\begin{fact}
$f$ is a ps-morphism.
\end{fact}

Let us recall that this implies that $f$ is injective

\medskip

Assume by contradiction that $f$ is not $k$-power-free.

We first make a crucial choice.

\medskip

\noindent
\textit{Choice 1}: let $W$ be a $k$-power-free word of
\textit{smallest} length such that $f(W)$ directly covers a $k$-power.

\medskip

Let $U$ be a word such that $U^k$ is directly covered by $f(W)$.

Since $\TSUA \subseteq \TSU$, $|W| \geq k+1$.  Let $(p_i, s_i, a_i,
W_i)_{i = 0, \ldots, k}$ be a decomposition of $U^k$ in $f(W)$. By
Lemma~\ref{lemmaSynchro}, this decomposition is not synchronized, that
is, $s_i \neq s_j$ and $p_i \neq p_j$ for all integers $i, j$ with $0 <
i < j < k$.

\medskip

Applying iteratively the Reduction Lemma~\ref{reduc} with the
deterministic rules chosen in Section~\ref{more}, we construct
some words $w$ and $u$ such that $(u^k, w)$ belongs to ${\rm Reduced}(U^k, W)$,
$|w| \leq |W|$ and $w \in \TSUC$.  We know that the decomposition of
$u^k$ in $f(w)$ is $(p_i, s_i, a_i, w_i)_{i = 0, \ldots, k}$ for some
words $(w_i)_{i = 0, \ldots, k}$.

Let us observe that since the
decomposition $(p_i, s_i, a_i, W_i)_{i = 0, \ldots, k}$ is not
synchronized, it follows the definition that $(p_i, s_i, a_i, w_i)_{i
= 0, \ldots, k}$ is also not synchronized.

Since $w \in \TSUC$ and $f(\TSUB)$ is $k$-power-free, we deduce that:

\begin{fact}
\label{fact2}
$w$ is not $k$-power-free.
\end{fact}

\noindent
\textit{Choice 2}:
let $v^k$ be a smallest $k$-power factor of $w$ ($v \neq \varepsilon$).

We denote $v_1, v_2$ words such that $w = v_1v^k v_2$.

\begin{fact}
\label{fact3}
No powers respectively of $f(v)$ and of $u$ have a common factor of
length greater than or equals to $|f(v)|+|u|-gcd(|f(v)|, |u|)$.
\end{fact}

This fact is a consequence of the following proposition which is a
corollary of the well-known Fine and Wilf's theorem (see
\cite{Lot1983,Lot2002} for instance).

\begin{prop}{\rm \cite{Ker1986}}
\label{Kera} Let $x$ and $y$ be two words. If a power of $x$ and a
power of $y$ have a common factor of length at least equal to
$|x|+|y|-gcd(|x|,|y|)$ then there exist two words $t_1$ and $t_2$
such that $x$ is a power of $t_1t_2$ and $y$ is a power of
$t_2t_1$ with $t_1t_2$ and $t_2t_1$ primitive words. Furthermore,
if $|x|>|y|$ then $x$ is not primitive.
\end{prop}

\begin{proof2}{Fact~\ref{fact3}}
Assume the opposite. By Proposition~\ref{Kera}, there exist two words
$t_1, t_2$ and two integers $n_1, n_2$ such that $f(v) =
(t_1t_2)^{n_1}$ and $u = (t_2t_1)^{n_2}$. Since $u \neq \varepsilon$
and $v \neq \varepsilon$, we have $t_1t_2 \neq \varepsilon$, $n_1 \geq
1$ and $n_2 \geq 1$. If $n_1 \geq 2$, $f(v^{\lceil k/2\rceil}) =
(t_1t_2)^{n_1\lceil k/2\rceil}$ contains the $k$-power
$(t_1t_2)^k$. Since $k \geq 3$, $\lceil k/2\rceil < k$, and so
$|v^{\lceil k/2\rceil}| < |v^k| \leq |w| \leq |W|$. By choice of $v$,
$v^{\lceil k/2\rceil}$ is $k$-power-free: this contradicts Choice~1 on
$W$. So $n_1 = 1$. We get $|u|=|f(v)^{n_2}| = n_2|f(v)|$ and so $|u| =
0 \mod L$.  For all integers $j$ between 1 and $k$, $|u s_j| =
|s_{j-1} f(w_j)p_js_j| = |s_{j-1} f(w_ja_j)|$, and so $|s_j| =
|s_{j-1}| \mod L$.  But for $j \geq 1$, $p_j \neq \varepsilon$, so
that $|s_j| < L$. It follows that $|s_j| = |s_{j-1}|$ for all $j \geq
2$.  This contradicts the fact that the decomposition of $u^k$ is not
synchronized.
\end{proof2}

\begin{fact}
\label{newFact4}
$|v| = 1$ and $k =3$
\end{fact}

The proof of this fact is made of three steps.

\medskip

\noindent
\textbf{Step~\ref{newFact4}.1}
If $|f(v)| \geq |u|$ then $|v| = 1$ and $k =3$.

\begin{proof}
Since $v \neq \varepsilon$, we can write $v=xv'=v''y$ for two letters
$x,y$ and two words $v',v''$.  Since $f(w)=f(v_1)f(v)^kf(v_2)=p u^k s$
with $|p| < L$ and $|s| < L$, the word $C = f(v'v^{k-2}v'')$ is a
common factor of $f(v^k)$ and $u^k$.  We have $|C| = |f(v)^k| -
|f(xy)| = kL|v| - 2L$.  When $|v|\geq 2$ or when $k \geq 4$, $|C| \geq
2L|v| = 2|f(v)| \geq |f(v)| + |u|$. So by Fact~\ref{fact3}, we cannot
have $|v| \geq 2$ or $k \geq 4$, that is (since $v \neq \varepsilon$),
we must have $|v| = 1$ and $k = 3$.
\end{proof}

\medskip

\noindent
\textbf{Step~\ref{newFact4}.2}
If $|f(v)| < |u|$ then $|v| = 1$.

\begin{proof}
Let us assume by contradiction that  $|v| \geq 2$.
There exist two letters
$x,y$ and a word $v'$ such that
$v=xv'y$.
Since $|f(v_1)f(v)^kf(v_2)|
=|f(w)|=|pu^k s| \geq  |u^k| > |f(v^k)|$, we
have $|v_1v_2|\geq 1$.

If $v_1 = \varepsilon$, we get $x = a_0$ and $f(x)=p_0s_0$.  Since
$|f(xv'yv^{k-1})| = |f(v)^k| <|u^k| \leq |pu^k|$, the word $C =
s_0f(v'yv^{k-1})$ is a prefix of $u^k$ and so a common factor of
$f(v)^k$ and $u^k$.  Let us recall that
$w \in \TSUC$.  This means in particular that $|w_1|_y \leq 1$ and so, since
$w = xw_1a_1\prod_{i = 2}^k w_ia_i$ starts with $xv'yxv'y$, $w_1$ is
a prefix of $v'yxv'$.  Consequently $|u| \leq |s_0f(v'yxv'y)|$.  It
follows that $|C| = |s_0f(v'yxv'y)| + (k-2)|f(v)| \geq |u| + |f(v)|$.  This
contradicts Fact~\ref{fact3}. So $v_1 \neq \varepsilon$.

Similarly we can prove that $v_2 \neq  \varepsilon$ and so $v^k = (xv'y)^k$ is a factor
of $w_1\prod_{i = 2}^{k} a_iw_i$.
Thus $f(v)^k$ is a common factor of $f(v)^k$ and $u^k$.
Since $w \in \TSUC$, we have
$|w_i|_x \leq 1$ and
$|w_i|_y \leq 1$ for all $1 \leq i \leq k$.
This implies that $|xv'yxv'y| \geq |w_i|+2$ for all $1 \leq i \leq k$ and thus
$|f(v)^2| = |f(xv'yxv'y)| \geq |u|$.
Consequently $|f(v^3)| \geq |f(v)| + |u|$, and once again we have a contradiction
with Fact~\ref{fact3}.
\end{proof}

\medskip

\noindent
\textbf{Step~\ref{newFact4}.3}
If $|f(v)| < |u|$ then $k = 3$.

\begin{proof}
By the previous step, we know that $|v| = 1$. So
$v=x$ for a letter $x$.  Since
$|f(v_1)f(x)^kf(v_2)|=|f(w)|=|pu^ks| \geq k |u| > k|f(x)|$, we
have $|v_1v_2|\geq 1$.

If $v_1 = \varepsilon$, $x^k$ is a prefix of $w$, $x = a_0$ and $f(x)
= p_0s_0$.  Since $|s_0f(x)^{k-1}| \leq |f(x)^k| < |u^k|$ and since
$s_0f(\prod_{i = 1}^k w_ia_i) = u^ks_k$, the word $s_0f(x)^{k-1}$ is a
prefix of $u^k$ and so a common factor of $f(v)^k$ and $u^k$.
Since $w \in \TSUC$, $|w_1|_x \leq 1$.  This implies $w_1 = \varepsilon$
or $w_1 = x$ and so $|u| = |s_0f(w_1)p_1| \leq |s_0f(xx)|$.  If $k
\geq 4$, $|s_0f(x)^{k-1}| \geq |s_0f(xx)|+|f(x)| \geq |f(v)| + |u|$: this contradicts
Fact~\ref{fact3}. So $k = 3$.

When $v_2 = \varepsilon$, symmetrically we can prove $k = 3$.

Now we consider the case where $v_1 \neq \varepsilon$ and $v_2 \neq
\varepsilon$. The word $f(x)^k$ is a common factor of $f(v)^k$ and
$u^k$.  Let us recall that $w = a_0 \prod_{i = 1}^k w_ia_i$, and
$|w_i|_x \leq 1$ for each integer $i$ with $1 \leq i \leq k$.  Since
here $x^k$ is a factor of $w_1 \prod_{i = 2}^{k-1}a_i w_{i+1}$, there
must exist an integer $i$, $1 \leq i \leq k$ such that $w_i =
x$. Thus $|u|+|f(x)| = |s_{i-1}f(w_i)p_i|+|f(x)|\leq |f(x^4)|$. This
contradicts Fact~\ref{fact3} when $k \geq 4$. So $k = 3$.
\end{proof}

\bigskip

We now make a break in the proof of the theorem to explain the
situation.  Up to now, we have proved this theorem when $k \geq 4$
showing that, when $f(\TSU)$ is $k$-power-free, there cannot exist
words like $w$ and $u$ such that $w \in \TSUC$ and $u^k$ has a
non-synchronized decomposition in $f(w)$. Example~\ref{examplek=3}
shows that this is possible when $k = 3$. Consequently when dealing
with Case $k = 3$ (and $|v| = 1$), we have to consider the
sequence of reductions of the couple $(U^k,W)$ into the couple $(u^k,w)$.
This will occur only in Cases~3 and 7 below. Actually
Example~\ref{examplek=3} belongs to Case~3.

\medskip

We now continue and end the proof of Theorem~\ref{TestSetUni}
treating Case $k = 3$ and $|v| = 1$.

Let us recall that $w = a_0w_1a_1w_2a_2w_3a_3 = v_1v^3v_2$. Since $|v|
= 1$, from now on, we replace the notation $v$ by $x$.  Since $w \in
\TSUC$, $|w_1|_x \leq 1$, $|w_2|_x \leq 1$, $|w_3|_x \leq 1$. Thus for
at least one integer $i$, $a_i = x$.  More precisely, we distinguish
nine cases depending on the relative position of $x^3$ with respect to
the $w_i$'s and to the $a_i$'s (see Figure~\ref{lesCas}: note that
Case~2 and Case~8 are split into two subcases).

\begin{figure}[ht]
\begin{center}
\begin{tabular}{|l|c|c|c|c|c|c|c|}
\hline
Case & $a_0$ & $w_1$ & $a_1$ & $w_2$ & $a_2$ & $w_3$ & $a_3$\\
\hline
1 & $x$ & $x$ & $x$ & & & & \\
\hline
2 & $x$ & $\varepsilon$ & $x$ & $x\ldots$ & & & \\
 & $x$ & $\varepsilon$ & $x$ & $\varepsilon$ & $x$ & & \\
\hline
3 & & \ldots $x$ & $x$ & $x\ldots$ & & & \\
\hline
4 & & \ldots $x$ & $x$ & $\varepsilon$ & $x$ & & \\
\hline
5 & & & $x$ & $x$ & $x$ & & \\
\hline
6 & & & $x$ & $\varepsilon$ & $x$ & $x\ldots$ & \\
\hline
7 & & & & $\ldots x$ & $x$ & $x\ldots$ & \\
\hline
8 & & & $x$ & $\varepsilon$ & $x$ & $\varepsilon$ & $x$ \\
  & & & & $\ldots x$ & $x$ & $\varepsilon$ & $x$ \\
\hline
9 & & & & & $x$ & $x$ & $x$\\
\hline
\end{tabular}
\end{center}
\caption{\label{lesCas}Nine cases}
\end{figure}

\begin{description}
\setlength\parsep{0pt}%
\setlength\itemsep{0pt}%
\item{Case 1:} $a_0 = x = w_1 = a_1$;

\item{Case 2:} $a_0 = x$, $w_1 = \varepsilon$, $a_1  = x$, $w_2a_2$
starts with $x$;

\item{Case 3:} $w_1$ ends with $x$, $a_1 = x$,
$w_2 \neq \varepsilon$,
$w_2$ starts with $x$;

\item{Case 4:} $w_1$ ends with $x$,
$a_1 = x$,
$w_2 = \varepsilon$,
$a_2 = x$;

\item{Case 5:} $a_1 = x = w_2 = a_2$;

\item{Case 6:} $a_1  = x$, $w_2 = \varepsilon$, $a_2 = x$,
$w_3$ starts with $x$;

\item{Case 7:} $w_2$ ends with $x$, $a_2 = x$,
$w_3$ starts with $x$;

\item{Case 8:} $a_1w_2$ ends with $x$, $a_2 = x = a_3$,
$w_3 = \varepsilon$;

\item{Case 9:} $a_2 = x = w_3 = a_3$.
\end{description}

Of course some cases are symmetric: Cases 1 and 9 (and Case 5 is very
close), Cases 2 and 8, Cases 3 and 7, Cases 4 and 6. In what follows
we prove that all cases are impossible
since they contradict previous facts or hypotheses. Firstly:

\begin{fact}
\label{fact7}
Cases 2, 4, 6 and 8 are not possible.
\end{fact}

Indeed in this cases, we can see that $u^3$ and $f(x)^3$ have a common
factor of length $|u|+|f(x)|$ (for instance in Case~2, the common factor
is $s_0p_1s_1x = us_1f(x)$): this contradicts Fact~\ref{fact3}.

\medskip

\begin{fact}
\label{fact8}
Case 1 is not possible.
\end{fact}

\begin{proof}
In this case, we have $f(x) = p_0s_0 = p_1s_1$ (see Figure~\ref{figCas1}).

\begin{figure}
\begin{center}
\epsf{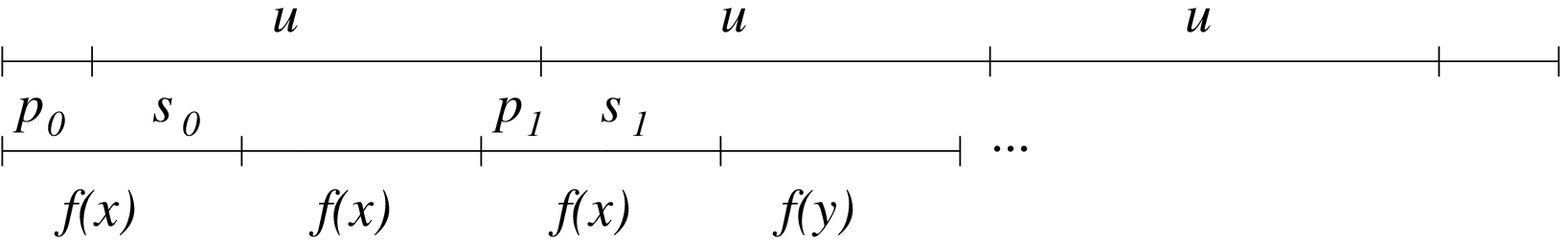}[xscale=5/8,yscale=5/8]
\end{center}
\def\figurename{Figure}
\def\captionseparator{~}
\caption{Case 1\label{figCas1}}
\end{figure}

Let $y$ be the first letter of $w_2a_2$ (that is, $w_2 \neq
\varepsilon$ and $y$ is the first letter of $w_2$, or, $w_2 =
\varepsilon$ and $a_2 = y$).  Assume $y = x$.  The word $u =
s_0f(x)p_1$ is a factor of $f(x^3)$.  Since $u^3$ is not synchronized
in $f(w)$, $|s_0| \neq |s_1|$.  If $|s_0| < |s_1|$, $s_0f(x)$ is a
prefix of $s_1f(x)$.  If $|s_1| < |s_0|$, $s_1f(x)$ is a prefix of
$s_0f(x)$.  In both cases $f(x)$ is an internal factor of $f(xx)$: By
Lemma~\ref{intfact}, $f(x)$ is not primitive. This implies that $f(xx)$
contains a 3-power, a contradiction with the 3-power-freeness of
$f(\TSU)$.

So $y \neq x$. We now consider two subcases.

\medskip

\noindent
\textit{Case 1.a}: $|s_0| > |s_1|$

Since $f$ is
uniform, $|f(x)| = |f(y)|$ and so $|s_0f(x)| > |s_1f(y)|$.  
In this case $u = s_0f(x)p_1$ starts with $s_1f(y)$.  
Let $v_3'$
be the word such that $s_0f(x) = s_1f(y)v_3'$.  We have $|v_3'| =
|s_0|-|s_1| \leq |s_0| \leq |f(a_0)| = |f(x)|$.  Thus $v_3'$ is a
suffix of $f(x)$.  Let $v_3$ be the word such that $f(x) = v_3v_3'$.
Since $s_0$ is a suffix of $f(x)$, $v_3'$ is a suffix of $s_0$. For
length reason, it follows that $f(y) = v_3'v_3$. If $|p_1| \geq |v_3|$
then, since $f(x) = p_1s_1 = v_3v_3'$, $v_3$ is a prefix of $p_1$ and,
since $f$ is injective, $yy$ is a factor of $w_2$: this contradicts the
fact that $w \in \TSUC$.

So $|p_1| < |v_3|$ and consequently $|s_1| > |v_3'|$ (since $f(x) =
p_1s_1 = v_3v_3'$). We observe that $s_0 = s_1v_3'$ and $s_1$ is a
suffix of $s_0$ (remember $f(x) = p_0s_0$), that is $s_0 = s_1v_3' =
v_4 s_1$ for a word $v_4$.  Lemma~\ref{Lothaire} implies the existence
of words $\alpha$, $\beta$ and of an integer $r$ such that $s_1 =
(\alpha\beta)^r\alpha$, $v_3' = \beta\alpha$ and $s_0 =
(\alpha\beta)^{r+1}\alpha$ (and $v_4 = \alpha\beta$). Note that $r
\geq 1$ and $\alpha\beta \neq \varepsilon$ since $|s_1| > |v_3'| \neq 0$.
Thus the words $f(xy)$ contains the factor $s_0v_3' =
(\alpha\beta)^{r+2}\alpha$ which contains the 3-power
$(\alpha\beta)^3$. Since $xy \in \TSU$, this contradicts the
3-power-freeness of $f(\TSU)$.

\medskip

\noindent
\textit{Case 1.b}: $|s_0| < |s_1|$

We have $f(x) = p_0s_0 = p_1s_1$ and $u$ starts with both $s_0$ and
$s_1$.  Let $v_3$ and $v_4$ be the words such that $v_3s_0 = s_1 =
s_0v_4$.  Lemma~\ref{Lothaire} implies the existence of words $\alpha,
\beta$ and of an integer $r$ such that $v_3 = \alpha\beta$, $s_0 =
(\alpha\beta)^r\alpha$, $v_4 = \beta\alpha$.  Since $u$ starts both
with $s_0f(x)$ and with $s_1$, the word $v_4$ is a prefix of $f(x)$. Thus $f(xx)$ contains the factor $s_1v_4 = (\alpha\beta)^{r+2}\alpha$.
Since $xx \in \TSU$ and
$f(\TSU)$ is 3-power-free, we have $r = 0$, that is $s_1 = s_0\beta s_0$.

If $|f(x)| \geq |\beta s_0| + |s_1|$, then $f(x) = v_4 t s_1 = \beta s_0 t s_0
\beta s_0$ for a word $t$ and $p_1 = \beta s_0 t$: $u = s_0 f(x) p_1 =
s_0 \beta s_0 t s_0 \beta s_0 \beta s_0 t$. Since the word $uu$ starts
with $s_1f(y) = s_0\beta s_0 f(y)$ and since $|f(y)| = |f(x)|$, we
have $f(y) = t s_0 \beta s_0 \beta s_0$. It follows that $f(yx)$
contains the 3-power $(s_0\beta)^3$: this contradicts the
3-power-freeness of $f(\TSU)$.

So $|f(x)| < |\beta s_0| + |s_1|$.  Since $u$ starts with $s_0f(x)$
and with $s_1 = s_0\beta s_0$, there exists a word $v_5$ such that
$f(x) = (\beta s_0) v_5$.  Let us recall that $f(x) = p_1s_0(\beta
s_0)$.  Lemma~\ref{Lothaire} implies the existence of some words
$\gamma, \delta$ and of an integer $s$ such that $p_1 s_0 = \gamma
\delta$, $\beta s_0 = (\gamma \delta)^s \gamma$ and $v_5 = \delta
\gamma$.  The word $uu$ starts both with $s_0 f(x) = s_0 \beta s_0v_5$
and with $s_1f(y) = s_0 \beta s_0 f(y)$. Consequently $v_5$ is a
prefix of $f(y)$ and $f(xy)$ starts with
$(\gamma\delta)^{s+2}\gamma$. Since $xy \in \TSU$ and $f(\TSU)$ is
3-power-free, $s = 0$: $\gamma = \beta s_0$, $p_1s_0 = \beta s_0
\delta$, $v_5 = \delta \beta s_0$ and
$f(x) = \beta s_0 \delta \beta s_0$.
Let us recall that
$|p_1|+|s_1| = |f(x)| < |\beta s_0| + |s_1|$, that is $|p_1| < |\beta
s_0|$. Consequently from $p_1 s_0 = \beta s_0 \delta$, we deduce that
$\delta$ is a suffix of $s_0$. Thus $f(xxy)$ contains the factor
$s_0f(x)v_5$ that ends with $(\delta\beta s_0)^3$. Since $x \neq y$,
$xxy \in \TSU$. We have a contradiction with the 3-power-freeness of
$f(\TSU)$. This ends the proof of impossibility of Case~1.
\end{proof}

\medskip

As already said, Case~9 is symmetric to Case~1. Moreover Case~5 can be
treated as previously. Hence

\begin{fact}
\label{fact9} Cases 5 and 9 are not possible
\end{fact}

We end the proof of Theorem~\ref{TestSetUni} with the proof of the 
final following case:

\begin{fact}
Cases 3 and 7 are not possible.
\end{fact}

\begin{proof}
Since Cases~3 and 7 are symmetric, from now on
we only consider Case~3.
Let us consider the sequence of words obtained by successive
reductions of $W$ leading to $w$. More precisely, let $(\nu_i,
\sigma_i)_{1 \leq i \leq m}$ be the couple of words such that $(\nu_1,
\sigma_1) = (U, W)$, $(\nu_m, \sigma_m) = (u, w)$ and for each $i$, $1
\leq i < m$, $(\nu_{i+1}^k, \sigma_{i+1})$
is the word in ${\rm Reduced}(\nu_i^k, \sigma_i)$ obtained by applying
Lemma~\ref{reduc} with the additional Reduction Rules chosen in
Section~\ref{more}.  By the reduction process, we know that each
one of the $k$-powers $\nu_j^k$ ($1 \leq j \leq m$) has a $(p_i, s_i,
a_i, w_{i,j})_{0 \leq i \leq 3}$-decomposition in $f(\sigma_i)$ for
some words $(w_{i,j})_{1 \leq i \leq 3}$.

By hypotheses of Case~3, $\sigma_m = w$ contains the 3-power $x$
centered in $a_1$. We mean more precisely that $w_1 = w_{m,1}$ ends
with $x$ (and so is not the empty word), $a_1 = a_{m,1} = x$ and $w_2 =
w_{m,2}$ starts with $x$ (and is also not the empty word). On other part,
$\sigma_1 = W$ is 3-power-free. Thus there exists an integer $q$ with
$1 \leq q < m$ such that $\sigma_q$ does not contains $xxx$ centered
in $a_1$ where'as $\sigma_j$ contains $xxx$ centered in $a_1$ for all
$j$ such that $q+1 \leq j \leq m$. To simplify temporarily the
notation, we set $W_1 = \sigma_q$, $W_2 = \sigma_{q+1}$, $U_1 = \nu_q$
and $U_2 = \nu_{q+1}$.

By the reduction process, there exist words $(x_i,y_i,z_i)_{1 \leq i \leq
3}$ (set also $x_0 = y_0 = z_0 = \varepsilon$) such that $U_1^3$ has a
$(p_i, s_i, a_i, x_iy_iz_i)_{i = 0, \ldots, 3}$-decomposition in
$f(W_1)$ and $U_2^3$ has a $(p_i, s_i, a_i, x_iz_i)_{i = 0, \ldots,
3}$-decomposition in $f(W_2)$. Since $W_2$ is obtained from $W_1$ by
the Reduction Lemma~\ref{reduc} there exist an integer $1 \leq \ell
\leq 3$ and a letter $a$ in $A$ such that both $x_{\ell}$ and
$y_{\ell}$ end with $a$ and $|s_{\ell -1}f(x_{\ell})|-|f(a)| <
|s_{i-1}f(x_i)| \leq |s_{\ell -1}f(x_{\ell})|$ and $|y_i|=|y_{\ell}|$.
By the Reduction Rule~\ref{Newspered2}, $|x_{\ell}|_a=|y_{\ell}|_a=1$.

Finally let us stress that by definition of $W_1$ and $W_2$, we assume
that $x_1z_1$ ends with $x$, $x_2z_2$ starts with $x$ and that
either $x_1y_1z_1$ does not end with $x$ or $x_2y_2z_2$ does not start
with $x$.  We end in two steps showing first that $x_1y_1z_1$ must end with $x$, and second that $x_2y_2z_2$ must start with $x$: This contradicts the previous sentence.

\medskip

\noindent
\textbf{Step~1: $x_1y_1z_1$ must end with $x$}

Assume by contradiction that $x_1y_1z_1$ does not end with $x$.  Since
$x_1z_1$ ends with $x$, we have $z_1 = \varepsilon$ and $y_1$ ends
with $b \neq x$ (since $x_{\ell}$ and $y_{\ell}$ ends with the same
letter, it also means that $l \neq 1$). By Remark~\ref{spere}(\ref{spered5}),
$z_{\ell}=\varepsilon$ and $|p_{\ell}| \leq |p_1|$.  Thus
$U_1=s_0f(x_1y_1)p_1=s_{\ell -1}f(x_{\ell}y_{\ell})p_{\ell}$ with
$|y_1|=|y_{\ell}|$, $x_1(=x_1z_1)$ ends with $x$ and both $x_{\ell}$
and $y_{\ell}$ end with $a$.  Let $c$ be the first letter of $y_1$
(see figure~\ref{L1a}).
\begin{figure}[ht]
\begin{center}
\ \\[0pt]
\epsfig{file=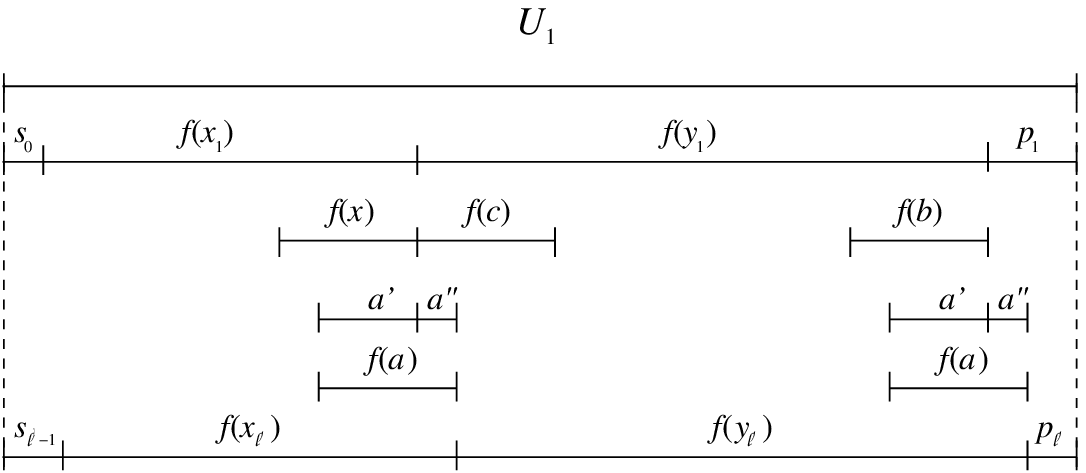}
\end{center}
\def\figurename{Figure}
\def\captionseparator{~}
\caption{\label{L1a}}
\end{figure}

Let $a''$ be the suffix of $f(a)$ such that $p_1=a''p_{\ell}$ and let
$a'$ be the prefix of $f(a)$ such that $f(a)=a'a''$. Since $f(b)p_1$
and $f(a)p_{\ell}$ are both suffixes of $U_1$, we get that $f(b)$ ends
with $a'$.  Since $|f(y_1)p_1|=|f(y_{\ell})a''p_{\ell}|$, we get that
$|s_0f(x_1)a''|= |s_0f(x_1y_1)p_1|+|a''|-|f(y_1)p_1| =|s_{\ell -1}f(x_{\ell})|$.
So $f(x)$ ends with $a'$
and $f(c)$ starts with $a''$. Since $p_1$ and so $a''$ are prefixes of
$f(x)$, by a length criterion, it follows that $f(x)=a''a'$.

If $c \neq x$, $bx^2c$ is 3-power-free and $f(bx^2c)$ contains the
3-power $(a'a'')^3$: this contradicts the 3-power-freeness of
$f(\TSU)$.

Thus $c = x$. If $|x_{\ell}| \geq 2$, let $e$ be the letter such that
$x_{\ell}$ ends with $ea$. Since $y_1$ contains $b$ and $c$ with $b
\neq x = c$, we have $|y_{\ell}| = |y_1| \geq 2$.  Let $d$ be the
first letter of $y_{\ell}$. We have $d \neq a$ and $e \neq a$ since
$|x_{\ell}|_a=|y_{\ell}|_a=1$.  Since
$f(y_1)p_1=a''f(y_{\ell})p_{\ell}$ and since $f(y_1)$ starts with
$f(x)=a''a'$, we get that $f(d)$ starts with $a'$.  Since
$s_0f(x_1)a''=s_{\ell-1}f(x_{\ell})$ and since $f(x_1)$ ends with
$f(x)$, we get that $f(e)$ ends with $a''$. It follows that $f(ea^2d)$
contains $(a''a')^3$ although $ea^2d$ is a 3-power free word: this
contradicts the 3-power-freeness of $f(\TSU)$.

Thus $ c = x$ and $|x_{\ell}|=1$. Consequently $x_{\ell}=a$.  Since
$|s_{\ell-1}f(x_\ell)| - |f(a)| < |s_0f(x_1)| \leq |s_{\ell-1}f(x_\ell)|$ and
since $x_1$ ends with $x$, we have $x_1 = x$.  Thus $U_2 =
s_0f(x)p_1$.  Since $|u| \geq |s_0f(x)p_1|$, we deduce that $U_2 = u$
and $W_2 = w$.  It follows that $u = s_0f(x)p_1$.

Let us recall that moreover $f(x) = p_1s_1$ and $s_1f(x)$ is a prefix
of $u$.  If $|s_0| < |s_1|$ then $f(x)$ is an internal factor of
$f(xx)$ and (by Lemma~\ref{intfact}) $f(x^2)$ contains a 3-power: this
contradicts the 3-power-freeness of $f(\TSU)$.  Thus $|s_0| \geq
|s_1|$. Let $s_0''$ be the suffix of $s_0$ and $p_2'$ be the word such
that $s_0''f(x)= f(x)p_2'$ and $s_1f(x)p_2'$ is a prefix of $u$. By
Lemma~\ref{Lothaire}, there exist two words $\alpha$ and $\beta$ such
that $s_0''=\alpha\beta (\neq \varepsilon)$, $p_2'=\beta\alpha$ and
$f(x)=(\alpha\beta)^r\alpha$ for an integer $r$.  We have $|s_0''| =
|s_0|-|s_1| \leq |f(a_0)|$. If $|s_0''| = |f(a_0)|$, then
$s_0=f(a_0)$ and $s_1 = \varepsilon$:
this contradicts the fact that
$u^k$ is not synchronized in $f(w)$.  Thus $|s_0''| < |f(a_0)| =
|f(x)|$. Consequently $r \geq 1$.  Let $\gamma$ be the letter such
that $x\gamma$ is a prefix of $w_2a_2$: $p_2'$ is a prefix of
$f(\gamma)$.  By Fact~\ref{fact3}, no powers respectively of $f(x)$
and of $u^3$ have a common factor of length greater than
$|f(x)|+|u|$. Hence $a_0 \neq x$.  But then
$a_0x\gamma$ is 3-power-free where'as $f(a_0x\gamma)$ contains
$(\alpha\beta)^{r+2}$: this contradicts the 3-power-freeness of
$f(\TSU)$, and so $x_1y_1z_1$ must end with $x$.

\medskip

\noindent
\textbf{Step 2: $x_2y_2z_2$ must start with $x$}

From what precedes, we know now that it remains to consider the case where
$x_2y_2z_2$ does not start with $x$. We will show that this assumption
leads to a final contradiction.

Since $x_2z_2$ starts with $x$, we have $x_2 = \varepsilon$ and $y_2$
starts with $b \neq x$. By Remark~\ref{spere}(\ref{spered3}),
$x_{\ell}=a$ and $|s_{\ell-1}| < |s_1|$ (and so $l \neq 2$).  Thus $U_1
=s_1f(y_2z_2)p_2=s_{\ell -1}f(ay_{\ell}z_{\ell})p_{\ell}$ with
$|y_2|=|y_{\ell}|$, $z_2=x_2z_2$ starts with $x$ and $y_{\ell}$ ends
with $a$.  Let $c$ be the last letter of $y_2$ (see
figure~\ref{L2a}).
\begin{figure}[ht]
\begin{center}
\ \\[0pt]
\epsfig{file=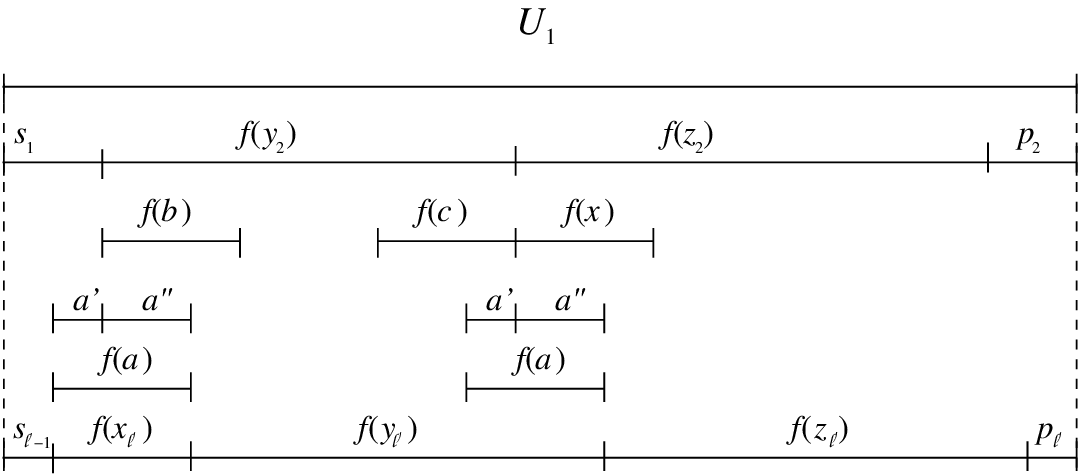}
\end{center}
\def\figurename{Figure}
\def\captionseparator{~}
\caption{\label{L2a}}
\end{figure}

Let $a'$ be the prefix of $f(a)$ such that $s_1=s_{\ell-1}a'$ and let
$a''$ be the suffix of $f(a)$ such that $f(a)=a'a''$.  Since
$s_{\ell-1}f(a)$ and $s_1f(b)$ are both prefixes of $U_1$, the word
$f(b)$ starts with $a''$.  Since
$|s_1f(y_2)|=|s_{\ell-1}a'f(y_{\ell})|=|s_{\ell-1}f(ay_{\ell})|-|a''|
=|U_1|-|f(z_{\ell})p_{\ell}a''|$, we have
$|f(z_2)p_2|=|U_1|-|s_1f(y_2)|=|a''f(z_{\ell})p_{\ell}|$.  Since
$f(z_2)p_2$ and $a''f(z_{\ell}p_{\ell})$ are both suffixes of $U_1$,
it follows that $f(z_2)p_2=a''f(z_{\ell})p_{\ell}$ and we get that
$f(x)$ starts with $a''$ and $f(c)$ ends with $a'$. Since $a'$ is a
suffix of $s_1$ and so of $f(x)$, by a length criterion, we get
$f(x)=a''a'$.

If $c \neq x$, $cx^2b$ is 3-power-free and $f(cx^2b)$ contains
$(a'a'')^3$: this contradicts the 3-power-freeness of $f(\TSU)$.

Thus $c = x$ and $y_2$ contains two different letters $b$ and $x$.
We get $|y_{\ell}|=|y_2|\geq 2$. Let $d$ be the letter such that
$y_{\ell}$ ends with $da$.  Since $|y_{\ell}|_a=1$, we have $d \neq
a$.  Since $f(x)a''$ and $f(da)$ are both suffixes of $f(y_{\ell})$,
the word $f(d)$ ends with $a''$.
Since $x_2z_2 = z_2$ starts with
$x$, $|z_2|_x \neq 0$.  Let $z_2'$ and $z_2''$ be the words such that
$z_2=z_2'xz_2''$ with $|z_2''|_x=0$.  Let $z_{\ell}'$ be the word and
$e$ be the letter such that $z_{\ell}'e$ is the prefix of
$z_{\ell}a_{\ell}$ verifying $|s_{\ell -1}f(ay_{\ell}z_{\ell}')|
<|s_1f(y_2z_2'x)| \leq |s_{\ell -1}f(ay_{\ell}z_{\ell}'e)|$.  Let us
recall that $s_1 = s_{\ell-1}a'$ and so $|s_1| = |s_{\ell-1}a'|$.  Moreover
$s_1f(y_2z_2'x)$ and $s_{\ell -1}f(ay_{\ell}z_{\ell}')$ are both
prefixes of $U_1$, and $|s_1a''| = |s_{l-1}| + |f(x)| = |s_{l-1}| \mod L$. Thus $s_1f(y_2z_2'x)=s_1f(y_2z_2')a''a'=s_{\ell
-1}f(ay_{\ell}z_{\ell}')a'$.  It follows that $f(e)$ starts with $a'$.
If $e \neq a$, $da^2e$ is 3-power-free and $f(da^2e)$ contains
$(a''a')^3$: this contradicts the 3-power-freeness of $f(\TSU)$.

Thus $e = a$.  Assume $|z_\ell'e| \leq |z_\ell|$.  Let us recall that
the reductions are assumed to be made under two rules.  The second
Reduction Rule implies that, having made a reduction with, in
Lemma~\ref{reduc}, an integer $\ell$ and a letter $a$, then if $|z_{\ell}|_a
\neq 0$, the next $|z_{\ell}|_a$ reductions are made with the same integer
$l$ and the same letter $a$. Thus here the words $\sigma_{q+2}$,
\ldots, $\sigma_{q+1+|z_\ell'a|_a}$ exist and are obtained using, in
Lemma~\ref{reduc}, the same integer $\ell$ and the same letter $a$
than the ones used to reduce $\sigma_q = W_1$ into $\sigma_{q+1} =
W_2$.  Moreover $\nu_{q+1+|z_\ell'a|_a} = s_1f(z_2'')p_2$.  Since
$|\nu_{q+1+|z_\ell'a|_a}| \geq |\nu_m| \geq |s_1f(x)p_2|$, we have
$z_2'' \neq \varepsilon$. But since $|z_2''|_x=0$, we have a contradiction with the
fact that $w$ contains $xxx$ centered in $a_1$.

Thus $|z_\ell'e| > |z_\ell|$, that is, $z_\ell = z_\ell'$, $e = a_\ell
(=a)$.  It follows that $z_2'' = \varepsilon$. Since $\sigma_j$
contains $xxx$ centered in $a_1$ for all $q+1 \leq j \leq m$, we must
have $w_2 = x$ and $u = s_1f(x)p_2$. Let us recall that moreover $f(x)
= p_1s_1$ and $f(x)p_1$ is a suffix of $u$.  If $|p_1| > |p_2|$ then
$f(x)$ is an internal factor of $f(xx)$ and (by Lemma~\ref{intfact})
$f(x^2)$ contains a 3-power: this contradicts the 3-power-freeness of
$f(\TSU)$. Since the decomposition is not synchronized, we have $|p_1| \neq |p_2|$.
Thus $|p_1| < |p_2|$. Let $p_2''$ be the prefix of $p_2$
and $s_0'$ be the word such that $p_2 = p_2''p_1$, $f(x)p_2''=
s_0'f(x)$ and $s_0'f(x)p_1$ is a suffix of $u$. By
Lemma~\ref{Lothaire}, there exist two words $\alpha$ and $\beta$ such
that $s_0'=\alpha\beta (\neq \varepsilon)$, $p_2''=\beta\alpha$ and
$f(x)=(\alpha\beta)^r\alpha$ for an integer $r$. Since $|s_0'| =
|p_2|-|p_1| < |f(x)|$ (remember $|p_1| \neq 0$), we have $r \geq 1$.
Let $\gamma$ be the letter such that $\gamma x$ is a suffix of
$a_0w_1$: $s_0'$ is a suffix of $f(\gamma)$.  By Fact~\ref{fact3}, no
power respectively of $f(x)$ and of $u^3$ have a common factor of
length greater than $|f(x)|+|u|$. Hence $a_2 \neq x$.
But then $\gamma xa_2$ is 3-power-free where'as $f(\gamma x a_2)$
contains $(\alpha\beta)^{r+2}$: this contradicts the 3-power-freeness
of $f(\TSU)$. This is a final contradiction proving that Case~3 is not
possible. So consequently Theorem~\ref{TestSetUni} holds.
\end{proof}


\section{Conclusion}

Theorem~\ref{TestSetUni} and Corollary~\ref{Bound} lead to some
natural questions: is $\TSU$ the smaller test-set?  Is
the bound $b_{k,A} = k \times \card{A} + k +1$ optimal?  The answer to
these questions are negative at least in most of the previously known
cases. As already mentioned in the introduction, M.~Leconte
\cite{Lec1985} has previously got a test-set when $\card{A} = 3$. He
proved \cite{Lec1985}~: a uniform morphism $f$ defined on a three-letter
alphabet is $k$-power-free ($k \geq 3$) if and only if the images of
all $k$-power-free words of length at most $3k+5$ are $k$-power-free. We
observe that in case $k = 3$, we obtain a better bound than
M.~Leconte. But in all other cases, the bound of M.~Leconte is better than our.
Another result shows the non-optimality of our bound $b_{k,A}$.  When
$\card{A} = 2$ (and $k \geq 3$), V.~Ker\"anen proved: a uniform and primitive
morphism defined on a two-letter alphabet is $k$-power-free if and
only if the images of length at most 4 are $k$-power-free. This bound
in this result does not depend on the value of $k$ and is far better
than our general bound $b_{k,\{a,b\}} \geq b_{3,\{a,b\}} = 13$.

\medskip

To end, let us mention further works.  In this paper, we propose a new
technic to tackle the decidability of $k$-power-freeness of uniform
morphisms. We are now looking to extension of this technic to the
decidability of $k$-power-freeness of arbitrary morphisms.



\end{document}